
\def\bold#1{\setbox0=\hbox{$#1$}%
\kern-.025em\copy0\kern-\wd0
\kern.05em\copy0\kern-\wd0
\kern-.025em\raise.0433em\box0 }
\def\SCIPP{\centerline {\it Santa Cruz Institute for Particle Physics}
\centerline{\it University of California, Santa Cruz, CA 95064}}
\def\ie{{\it i.e.}}
\def\hc{H^\pm}
\def\hl{h^0}
\def\hh{H^0}
\def\ha{A^0}
\def\call{{\cal L}}
\def\calo{{\cal O}}
\def\msusy{M_{SUSY}}
\def\msusyy{M^2_{SUSY}}
\Pubnum={SCIPP 92/29}
\date={July, 1992}
\pubtype{}
\titlepage
\vskip 4cm
\title{\bf Constraints from Global
 Symmetries on  Radiative Corrections
 to the Higgs Sector}
 \author{Howard E. Haber and Alex Pomarol}
\vskip .1in
\SCIPP
\vskip .2in

\vfill
\centerline{\bf Abstract}

We discuss the implications of global symmetries on the radiative
corrections to the Higgs sector.
We focus on two examples:
 the charged Higgs mass in the minimal supersymmetric model
 and the
Higgs couplings to vector boson pairs.
In the first case, we find that in the absence of squark mixing
a global
SU(2)$\times$SU(2) symmetry protects
the charged Higgs mass from corrections of $\calo (g^2m^4_t/m^2_W)$.
In the second case, it is the {\it custodial} symmetry which
plays an analogous role in constraining the fermion-mass dependence
of the radiative corrections.
\vskip .1in

\vfill
\endpage

\noindent{\bf 1. Introduction}

Global symmetries play an important role in analyzing the radiative
corrections of the tree-level parameters of a theory.
Often, a theory will possess a ``natural'' tree-level relation -- \ie,
 a relation among tree-level parameters which is attributable to some
  underlying symmetry. In this case, radiative corrections to this
 relation must be finite;
  moreover, the nature of the underlying symmetry can
provide information of the order of magnitude of these corrections.
As an example, in the Standard Model (SM)
\REF\cust{S. Weinberg, Phys. Rev. {\bf D19} (1979) 1277;
L. Susskind, Phys. Rev. {\bf D20} (1979) 2619;
P. Sikivie \etal, Nucl. Phys. {\bf B173} (1980) 189.}
the so-called global {\it custodial} SU(2) symmetry\refmark\cust\
 plays a crucial role in the analysis
of the radiative corrections to the $\rho$-parameter.
One of the most important implications of this global SU(2)
symmetry is the
\REF\screen{M. Veltman, Acta Phys. Pol. {\bf B8} (1977) 475; Phys. Lett.
{\bf B70} (1977) 253;
M.B. Einhorn and J. Wudka, Phys. Rev. {\bf D39} (1989) 2758.}
screening theorem of the Higgs boson\refmark\screen.

The purpose of this paper is to make use of global symmetries in
 the analysis of the radiative corrections to the Higgs sector.
The study of such radiative corrections
 in the minimal supersymmetric model (MSSM)
 has recently
 received much attention. One-loop effects have been found which
 significantly modify the
\REF\corr{H.E. Haber and R. Hempfling, Phys. Rev. Lett. {\bf 66}
(1991) 1815; SCIPP-91/33 (1992);
Y. Okada, M. Yamaguchi and T. Yanagida,
 Prog. Theor. Phys. {\bf 85} (1991) 1;
 Phys. Lett. {\bf B262} (1991) 54;
J. Ellis, G. Ridolfi and F. Zwirner,  Phys. Lett.
{\bf B257} (1991) 83;
 Phys. Lett. {\bf B262} (1991) 477;
R. Barbieri and M. Frigeni,  Phys. Lett. {\bf B258} (1991) 395;
R. Barbieri, M. Frigeni, F. Caravaglios,  Phys. Lett. {\bf B258}
 (1991) 167;
A. Yamada,  Phys. Lett. {\bf B263} (1991) 233;
D.M. Pierce, A. Papadopoulos and S.B. Johnson, Phys. Rev. Lett. {\bf
68} (1992) 3678;
A. Brignole, Phys. Lett. {\bf B281} (1992) 284.}
\REF\corrb{A. Brignole, J. Ellis, G. Ridolfi and F. Zwirner,
 Phys. Lett. {\bf B271} (1991) 123 [E: {\bf B273} (1991) 550];
M.A. D\'\i az and H.E. Haber, Phys. Rev. {\bf D45} (1990) 4246;
A. Brignole, Phys. Lett. {\bf B277} (1992) 313.}
tree-level predictions  for the Higgs masses
of the MSSM and give rise to
important phenomenological consequences\refmark{\corr,\corrb}.
For the light neutral  CP-even
 Higgs mass, radiative corrections
involving loop contributions from top quarks and their supersymmetric
partners
 induce a substantial squared mass shift
 of $\calo (g^2m^4_t/m^2_W)$.
However, for the charged Higgs squared mass, the radiative
corrections are not so important because (in the absence of squark
mixing) only one-loop corrections
of $\calo (g^2m^2_t)$ are induced.
In this paper, we shall show that these results follow easily by
studying the implications of the underlying global symmetries of the
Higgs potential.

In section 2, we make use of the global symmetries of the Higgs
potential to analyze the radiative corrections to the charged
Higgs mass in the MSSM. In particular, we will see that due to
  an approximate extended {\it custodial}
 symmetry of the Higgs potential, radiative corrections
 of ${\cal O}(g^2m^4_t/m^2_W)$ never arise.
In section 3, we study in a similar way
the one-loop effects to the
couplings of the Higgs bosons
to a pair of vector bosons. We shall demonstrate that the
{\it custodial} SU(2) symmetry plays a similar role to that in the
radiative corrections to the
$\rho$-parameter.
\bigskip
\noindent {\bf 2. Radiative corrections to the charged Higgs mass}

One of the relations that supersymmetry (SUSY)
imposes on the Higgs potential is the mass
\REF\hunter{See, for example
J.F. Gunion, G.L. Kane, H.E. Haber and S. Dawson,
 {\it The Higgs Hunter's Guide} (Addison-Wesley Publishing Company,
  Reading, MA, 1990) for a comprehensive review and a guide
 to the literature.}
 sum-rule\refmark\hunter
$$m^2_{\hc}=m^2_{\ha}+m^2_W\, .\eqn\sumrule$$
Because SUSY is not an exact symmetry of nature, eq.~\sumrule\
 only holds at tree-level
  and is modified by radiative corrections.
On dimensional grounds, one might naively
 expect that the radiative corrections to
eq.~\sumrule\ should depend quadratically on some large mass scale in
the problem. Specifically,
the largest contribution expected would come from loops
of superpartners whose masses are of the order of
the SUSY breaking scale,
$\msusy$. However,
such contributions are certainly absent
\REF\added{J.F. Gunion and A. Turski, Phys. Rev. {\bf D40} (1989) 2325.}
at one-loop for physical observables\refmark\added.
Specifically,
all one-loop
   corrections that grow as $\msusyy$ can be absorbed
in the redefinition of the mass-squared parameters of the Higgs
potential. In contrast, whereas
 these mass-squared parameters are all independent,
 the scalar self-couplings are related by SUSY. Therefore,
 we do not have enough freedom to absorb all the effects of the
 superpartners.
Since these effects can only show up in dimensionless
parameters, these will depend at most logarithmically on $\msusy$.
Note that decoupling does not apply when the mass
of a heavy particle, $M$, can be made large by increasing a dimensionless
parameter ({\it e.g.}, the masses of the fermions and the Higgs boson
 in the SM).
In that case, one-loop corrections to eq.~\sumrule\ of
 $\calo (M^2)$ can show up.
\REF\topmas{F. Abe \etal\  (CDF collaboration), Phys. Rev. Lett. {\bf 68}
(1992) 447.}

The recent experimental result that $m_t>91$ GeV\refmark\topmas\
suggests that radiative corrections to eq.~\sumrule\  due to the loop
 contributions of top quarks and top squarks should be the dominant
corrections. Naively, one expects one-loop corrections of order
$h_t^2m^2_t\sim g^2m^4_t/m^2_W$ where $h_t$ is the top
Yukawa coupling. Nevertheless, explicit calculation shows
that in the absence of squark mixing, the leading radiative corrections
 are only of order $g^2m^2_t$\refmark\corrb.

To exlain this result, we first analyze the two-doublet Higgs potential
before imposing SUSY. Let $\Phi_1$ and $\Phi_2$ denote two
Higgs doublets with hypercharges $Y=1$. The most general
renormalizable and SU(2)$_L\times$U(1)$_Y$ gauge invariant Higgs
potential is given by
$$\eqalign{V(\Phi_1,\Phi_2)=&
m^2_1\Phi^\dagger_1\Phi_1+m^2_2\Phi^\dagger_2\Phi_2
-(m^2_{12}\Phi^\dagger_1\Phi_2+h.c.)
+\lambda_1(\Phi_1^\dagger\Phi_1)^2+
\lambda_2(\Phi_2^\dagger\Phi_2)^2\cr
+&\ \lambda_3(\Phi_1^\dagger\Phi_1)(\Phi_2^\dagger\Phi_2)
+\lambda_4(\Phi_1^\dagger\Phi_2)(\Phi_2^\dagger\Phi_1)
+\coeff{1}{2}\left[\lambda_5(\Phi_1^\dagger\Phi_2)^2+h.c.\right]
 \, ,}\eqn\potencial$$
 where a  discrete symmetry
 $\Phi_2\rightarrow -\Phi_2$ has been imposed on the
  dimension-four terms.
This discrete symmetry guarantees the absence of
flavor changing neutral
\REF\fcnc{S.L. Glashow and S. Weinberg, Phys.
Rev. {\bf D15} (1977) 1958.}
current\refmark\fcnc.
It will be convenient to represent $\Phi_1$  by a real four vector, \ie,
$$\Phi_1=\left(\matrix{\phi^+_1\cr \phi^0_1}\right)=
\left(\matrix{\phi_3+i\phi_4\cr \phi_1+i\phi_2}\right)\rightarrow
\bold\Phi_1=(\phi_1,\phi_2, \phi_3,\phi_4)\, .\eqn\alexa$$
Since the MSSM Higgs sector automatically conserves CP at tree-level,
we henceforth make this assumption.
The physical spectrum of the model
consists in two charged Higgs bosons ($H^\pm$) and
three neutral ones: two CP-even ($\hl$ and $\hh$) and one CP-odd
($\ha$). The masses of the $\ha$ and $\hc$ are related by
$$m^2_{\hc}=m^2_{\ha}+\coeff{2m^2_W}{g^2}\left(\lambda_5-\lambda_4\right)
\, .\eqn\masses$$

Consider the limit where $m_{12}=\lambda_4=\lambda_5=0$. In this limit
 the global symmetry of the
 Higgs potential of eq.~\potencial\   is enlarged
  to O(4)$_1\times$O(4)$_2$.
Here, we find convenient to choose the symmetry transformations such
that $\bold\Phi_1$
transforms as a 4-vector under both O(4)$_1$ and O(4)$_2$, whereas
$\bold\Phi_2$
transforms as a 4-vector under O(4)$_1$ and as a singlet under O(4)$_2$.
When the scalar fields develop vacuum expectation values (VEVs),
 $\VEV{\bold\Phi_i}=(v_i,0,0,0)$, the O(4)$_1\times$O(4)$_2$
symmetry breaks down to O(3)$_1\times$O(3)$_2$ which is locally isomorphic
to SU(2)$\times$SU(2). Three of the six Goldstone bosons produced can be
associated with the breakdown of SU(2)$_L\times$U(1)$_Y\rightarrow
$U(1)$_{EM}$; these will be ``eaten" when the $Z$ and $W^\pm$ bosons
acquire mass.  The other
three Goldstone bosons
are the $A^0$ and the $H^\pm$.
If the broken  symmetries corresponding to the $\ha$ and $\hc$ Goldstone
bosons are symmetries of the full theory (prior to symmetry breaking),
then it would follow that
$m_{\ha}=m_{\hc}=0$ to all orders in
perturbation theory. In general, this will not be the case, in which case
the $\ha$ and $H^\pm$ are
pseudo-Goldstone bosons
(\ie, they would acquire a calculable mass due to radiative corrections).

Consider first the coupling of Higgs bosons to third generation quarks.
In supersymmetric theories, the coupling quark doublets to Higgs
doublets is such that $\Phi_1$ couples exclusively to $b_R$ and
$\Phi_2$ couples exclusively to $t_R$. We assume this coupling pattern
in the following.
In the limit $h_b=0$,
$$\call_Y=-h_t\left(\bar t_L\ \bar b_L
\right)  i\tau_2\Phi^*_2 t_R+h.c.\eqn\alexb$$
Since the global symmetry of this term is SU(2)$_L\times$U(1)$_Y\times$
O(4)$_2$, $t$-loop radiative corrections will not
 induce a mass terms
for the $\ha$ and the $\hc$. In the case of the MSSM,
 SUSY  imposes the following condition on the parameters of the
  two Higgs doublet potential of eq.~\potencial\refmark\hunter
$$\lambda_1=\lambda_2=\coeff{1}{8}(g^2+g^{\prime 2})\, ,\ \ \
\lambda_3=\coeff{1}{4}(g^2-g^{\prime 2})\, ,\ \ \ \lambda_4=-\coeff{1}{2}
g^2\, ,\ \ \ \lambda_5=0\, .\eqn\alexd$$
According to the above argument, the $\ha$ and $\hc$ must be
massless to all orders of perturbation theory in the limit of
$m_{12}=g=0$. That is,
 $t$-loop corrections to the mass sum-rule [eq.~\sumrule]
 must go to zero in
this limit.
 It follows that corrections of
$\calo (h^2_tm^2_t)$ to eq.~\sumrule\ must cancel out. In fact, each
term in the one-loop
radiative  corrections to eq.~\sumrule\ must
 depend quadratically on either $m_{\ha}$, $m_W$ or $m_b$.

 However, in the SUSY model, we must also
  consider the  squark sector since
  radiative corrections of
$\calo (h^2_tm^2_t)$ can also arise from top squark loops. Assuming
that there is no $\tilde t_L-\tilde t_R$
mixing, we find in the limit of  $h_b=g=0$
$$\call_{stop}=\call(\Phi^\dagger_1\Phi_1,\Phi^\dagger_2
\Phi_2, \tilde Q^\dagger\tilde Q,
\tilde U^*\tilde U)+h^2_t\left|\tilde Q^\dagger i\tau_2
\Phi_2^*\right|^2\, ,\eqn\alexe$$
where
$$\tilde Q=\left(\matrix{\tilde t_L\cr \tilde b_L}\right)\ \ \
{\rm  and}\ \ \ \tilde U=\tilde t^*_R\, .\eqn\alexg$$
These terms are also SU(2)$_L\times$U(1)$_Y\times$O(4)$_2$ invariant
and, therefore,
corrections of $\calo (h^2_tm^2_t)$ cannot
 arise from this sector either.
Finally, if $\tilde t_L-\tilde t_R$ is present,
 we have new terms given by
$$\call_{mix}=-\mu h_t\tilde Q^\dagger(i\tau_2\Phi^*_1)
\tilde U^*+h_tA_U\tilde Q^\dagger(i\tau_2\Phi_2^*)
\tilde U^*+h.c.\eqn\sq$$
which are not invariant under the global O(4)$_2$ symmetry.
Thus, top squark
loops involving the interactions of eq.~\sq\
 can induce corrections to eq.~\sumrule\ of
$\calo (h^2_tm^2_t)$.\foot{The one-loop top squark
corrections for large $\msusy$ are in
fact of $\calo \left[\mu^2 h^2_t(m^2_{\tilde t_L}-m^2_{\tilde b_L})/
m^2_{\tilde t_L}\right]$. However, if $\mu$ and the diagonal
 soft-supersymmetry breaking squark masses are
of the same order, we have that
$\mu^2(m^2_{\tilde t_L}-m^2_{\tilde b_L})/
m^2_{\tilde t_L}\sim  m^2_t$ resulting in a top squark correction of
 $\calo (h^2_tm^2_t)$.}
Notice that in the limit $\mu=0$ the terms in eq.~\sq\
 restore the O(4)$_2$
symmetry and, although we still have a $\tilde t_L-\tilde t_R$ mixing
($A_U\not= 0$), no corrections of $\calo (h^2_tm^2_t)$ can arise.
This results are in agreement with the explicit one-loop radiatively
corrected charged Higgs mass obtained in the literature\refmark\corrb.

Let us now analyze the corrections to eq.~\sumrule\
 from other sectors of the
theory. First, we consider
 the two Higgs doublet potential [eq.~\potencial]
 before imposing SUSY, where now
 we  take the limit $\lambda_4=\lambda_5$. In this limit the Higgs
potential is only O(4)$_1$ invariant. After spontaneous symmetry
breaking (SSB), the residual symmetry,
 O(3)$_1\sim$SU(2), is the so-called
{\it custodial} symmetry\refmark\cust\
which is responsible for the
relation $m^2_W=m^2_Z\cos^2\theta_W$. Setting $\lambda_4=\lambda_5$
in eq.~\masses\ yields
$$m^2_{\hc}=m^2_{\ha}\, .\eqn\ac$$
Radiative corrections to this relation will only come from sectors
of the theory not invariant under the global
{\it custodial} symmetry.
The {\it custodial} SU(2)
symmetry is an approximate
symmetry of the minimal supersymmetric
Higgs potential ($\lambda_4=-g^2/2\sim \lambda_5=0$). In the
limit $g\rightarrow 0$,
 \ie, $\lambda_4=\lambda_5$, eq.~\ac\ must hold to all orders in the
Higgs self-interactions. We conclude that
 the only non-vanishing correction to eq.~\sumrule\
from Higgs self-interactions must be proportional to
$g^2$.

Finally, let us consider the Higgs-gauge boson interactions. They derive
 from  the scalar kinetic term
$$\call_{kin}=\sum^2_{i=1}\coeff{1}{2}\tr\left\{(D^\mu M_i)^\dagger
 (D_\mu M_i)\right\}\, ,\eqn\kin$$
where
$$D_\mu M_i=\partial_\mu M_i+\coeff{1}{2}
 ig\bold{\tau}\bold\cdot{\bf W_\mu}M_i-\coeff{1}{2}ig^\prime B_\mu
  M_i\tau_3\, ,\eqn\taut$$
and
$$M_i=\left(i\tau_2\Phi^*_i\
 \Phi_i\right)\equiv\left(\matrix{\phi^{0*}_i&
\phi^+_i\cr -\phi^-_i&\phi^0_i}\right)\, .\eqn\mi$$
In the limit $g^\prime=0$, the kinetic term is invariant under the
global SU(2)$_L\times$SU(2)$_R \sim$\break O(4)$_1$ transformation,
$$\eqalign{M_i&\rightarrow L\, M_iR^\dagger\, ,\cr
\bold{\tau}\bold\cdot {\bf W}
&\rightarrow L\,\bold{\tau}\bold\cdot{\bf W} L^\dagger\, .}
\eqn\trans$$
After the neutral Higgs fields acquire VEVs
the residual symmetry of eq.~\kin\ [for $g^\prime=0$]
is SU(2)$_{L+R}$ which is the {\it custodial} symmetry described above.
Therefore, corrections to eq.~\masses\ are expected to be of
 $\calo [m^2_W(\lambda_4-\lambda_5)]$ for small {\it custodial}
breaking.
  In the MSSM it means corrections to $m^2_{\hc}$ of
 $\calo(g^2m^2_W)$. When the factor U(1)$_Y$ is gauged,
the presence of $\tau_3$ in eq.~\taut\ explicitily breaks the
  {\it custodial} symmetry  and corrections to $m^2_{\hc}$ of
 $\calo(g^{\prime 2}m^2_H)$, where $m_H$ is the largest Higgs mass, can
be generated. However, in
the MSSM, the Higgs masses can only be made substantially larger than
$m_Z$ increasing the soft $m^2_{12}$  mass-squared
parameter.\foot{This is not the case  of  the SM or a
non-supersymmetric two Higgs
doublet model. In these cases, the Higgs masses can be made
large by increasing the self-couplings $\lambda_i$.}
Therefore, one-loop corrections of
 $\calo (g^{\prime 2}m^2_H)$ must cancel in the large $m_H$ limit.

We end this section
with  a comment concerning the natural relation given in eq.~\masses\
which relates Higgs masses and the combination
($\lambda_4-\lambda_5$) of Higgs self-couplings. In principle,
($\lambda_4-\lambda_5$) can be measured independently of the masses.
Then, one can discuss finite radiative corrections to eq.~\masses.
The analysis is identical to the one presented above in the case of the
MSSM. Speciflcally, one-loop corrections terms to eq.~\masses\ can be
of $\calo (g^2m^2_t)$ or $\calo [m^2_W(\lambda_4-\lambda_5)]$.
In particular, no $\calo (g^2m^4_t/m^2_W)$ corrections can be generated
at one-loop. Since these corrections arise from the violation
of {\it custodial} symmetry, the size of these corrections  can be
constrained by the $\rho$-parameter ($\rho\equiv m^2_W/m^2_Z\cos^2
\theta_W$)  whose deviation from 1 also reflects the  presence
of {\it custodial} symmetry violating terms. It is well known that
 the size of $m_t$ (or $h_t$) is limited via this constraint. However,
 the dependence of the $\rho$-parameter on
($\lambda_4-\lambda_5$) can only occur at the two-loop level and
probably cannot provide a useful constraint.
\bigskip
\noindent{\bf 3. One-loop effective $\bold{HVV}$ vertices}

The trilinear $HVV$ vertices, where $H$ refers generically to any
Higgs boson and $V$ to any vector boson, are of interest for the
phenomenology of the Higgs bosons.
The $HVV$ vertices can provide an important
 production mechanism for Higgs bosons at future colliders.
Furthermore,
 the decay $H\rightarrow VV$ can be used as a clear signature of
 the $H$.
In Higgs sectors consisting in only doublets, $HVV$
\REF\grif{J.A. Grifols and A. M\'endez, Phys. Rev. {\bf D22} (1980)
1725;
A.A. Iogansen, N.G. Ural'tsev and V.A. Khoze, Sov. J. Nucl. Phys.
{\bf 36} (1983) 717.}
vertices are absent
at tree-level for the CP-odd and charged Higgs bosons\refmark\grif.
This is  the primary reason  why the $\ha$ and $\hc$ may be
difficult
to find at future hadron colliders.
The one-loop induced $\hc W^\mp Z$ and $\ha VV$
vertices in the MSSM have
\REF\hwz{A. M\'endez and A. Pomarol, Nucl. Phys. {\bf B349} (1991) 369;
M. Capdequi Peyran\`ere, H.E. Haber and P. Irulegui, Phys. Rev.
{\bf D44} (1991) 191.}
\REF\avv{A. M\'endez and A. Pomarol, Phys. Lett. {\bf B272} (1991) 313.}
\REF\kao{J.F. Gunion, H.E. Haber and C. Kao, Preprint UCD-91-32,
SCIPP-91/44 and FSU-HEP-91/222 (1991), Phys. Rev. {\bf D46} (1992),
in press.}
been calculated
in ref.$\, $[\hwz] and refs.$\, $[\avv,\kao]  respectively.
The primary contributions to the respective amplitudes
 arise from a virtual heavy quark pair.
In the case of the
$\hc W^\mp Z$ vertex, the  contribution of a heavy
 quark doublet ($u$,$\, d$) grows quadratically with the quark mass for
$m_u\not= m_d$. However, this leading contribution
 vanishes exactly if the heavy quarks in the doublet
 are mass-degenerate.
Scalar and gauge bosons contributions are found to be rather small due
to large cancellations among different diagrams.
As we shall see, such results are a consequence of the global
 {\it custodial} symmetry.
For simplicity, we will consider a sector with only two Higgs doublets.
The analysis, however, can be easily generalize
 to multi-doublet models.

Let us begin by assuming  that the global SU(2)$_{L+R}$ symmetry
 defined by eqs.~\trans\ with $L=R$ is an exact symmetry of our theory
even after SSB. Let us also work in the limit $g^\prime=0$.
In this case, the most general form for the one-loop effective
$HVV$ vertices is given by
$$\call_{HVV}=\sum_j{\bf O}^{\mu\nu}_j
\sum^2_{i=1}\mu_{ij}\tr\left\{M_i\bold{\tau}\bold\cdot
 {\bf W_\mu}\bold{\tau}\bold\cdot {\bf W_\nu}\right\}+h.c.\, ,\eqn\hvv$$
where
${\bf O}^{\mu\nu}_j=\left( g^{\mu\nu},\ \partial^\mu\partial^\nu,
\ \partial^\nu\partial^\mu,
\ \epsilon^{\mu\nu\rho\sigma}\partial_\rho\partial_\sigma\right)$
 and $\mu_{ij}$ are  complex constants
  of dimension $d=4-dim\{{\bf O}^{\mu\nu}_j\}$.
Using eq.~\mi, eq.~\hvv\ can be written as
$$\call_{HVV}\propto\sum_j{\bf O}^{\mu\nu}_j
\sum^2_{i=1}{\rm Re}\,\mu_{ij}
\left[W^3_\mu W^3_\nu+W^+_\mu W^-_\nu\right]
{\rm Re}\phi_i^0\, .\eqn\new$$
It is then clear that only the CP-even
 fields $\hl$ and
$\hh$ couple to a pair of gauge bosons. Thus the $\ha VV$
 and $\hc VV$ vertices will only be generated if the
{\it custodial} symmetry is violated.

 Let us analyze the quark-Yukawa sector and, in particular, its
 {\it custodial} limit.
 In a general model with two Higgs doublets,
there are two  possible ways to couple the Higgs to the quarks
 in a manner
 consistent with the discrete symmetry $\Phi_2\rightarrow - \Phi_2$:

{{\bf Case I: }Quarks couple only to the first Higgs doublet  $\Phi_1$.}

{{\bf Case II: }$\Phi_2$ couples to
 $u_R$ and $\Phi_1$ couples to   $d_R$.}

\noindent In case I,  the quark-Yukawa interactions are SU(2)$_L\times
$SU(2)$_R$    invariant if $h_u=h_d\equiv h$,
$$\call_Y=-h\left(\bar u_L\ \bar d_L \right)M_1
\left(\matrix{u_R\cr d_R}\right)+h.c.\, ,\eqn\alexh$$
where the relevant transformation laws are
$$\eqalign{\Psi_L\equiv\left(\matrix{u_L\cr d_L}\right)&\rightarrow
L\Psi_L\, ,\cr
\Psi_R\equiv\left(\matrix{u_R\cr d_R}\right)&\rightarrow R\Psi_R\, ,}
\eqn\alexi$$
$$M_1\rightarrow L\, M_1R^\dagger\, .\eqn\alexj$$
When the neutral scalars
develop  VEVs, the symmetry is broken down to
 SU(2)$_{L+R}$. The {\it custodial} limit, therefore, corresponds
 to the limit $m_u=m_d$.
Thus, we shall
 need a large mass splitting within the quark doublet
to generate  $\ha VV$ and $\hc VV$ vertices that are phenomenologically
relevant.

We now turn to case II (which is the quark-Higgs interactions required
by the MSSM).
If we define the transformation law of the scalar fields according to
eq.~\trans, we find that the quark Yukawa sector is not
SU(2)$_L\times$SU(2)$_R$ invariant, even
 in the limit $h_u=h_d$.
 However, by making the following redefinition,
$$\eqalign{\Phi_1&\rightarrow \coeff{1}{h_d}\Phi_1\, ,\cr
\Phi_2&\rightarrow \coeff{1}{h_u}\Phi_2\, ,}\eqn\alexk$$
the quark Yukawa sector can be written by
$$\call_Y=-\left(\bar u_L\ \bar d_L \right)\left(i\tau_2\Phi^*_2\
\Phi_1\right)
\left(\matrix{u_R\cr d_R}\right)+h.c.\, ,\eqn\alexk$$
which is SU(2)$_L\times$SU(2)$_R$ invariant
 if the scalar fields transform as
$$M_{21}\equiv\left(i\tau_2\Phi^*_2\ \Phi_1\right)\rightarrow L\, M_{21}
R^\dagger\, .\eqn\transtwo$$
After SSB the quark mass term is given by
$$\call_m=-\left(\bar u_L\ \bar d_L \right)
\left(\matrix{h_uv_2&0\cr 0&h_dv_1}\right)
\left(\matrix{u_R\cr d_R}\right)+h.c.\eqn\alexl$$
This term is
 SU(2)$_{L+R}$ invariant only if $h_uv_2=h_dv_1$, \ie, $m_u=m_d$.
Then, the effective $HVV$ vertices in the SU(2)$_{L+R}$
{\it custodial} limit (\ie, take $L=R$) are given by
$$\eqalign{\call_{HVV}=&\sum_j\mu_j{\bf O}^{\mu\nu}_j
\tr\left\{M_{21}\bold{\tau}\bold\cdot
 {\bf W_\mu}\bold{\tau}\bold\cdot {\bf W_\nu}\right\}+h.c. \cr
\propto&\sum_j{\bf O}^{\mu\nu}_j
\left(W^3_\mu W^3_\nu+W^+_\mu W^-_\nu\right)\left[{\rm Im}\, \mu_j
{\rm Im}(\phi_2^0-\phi^0_1)+{\rm Re}\, \mu_j
{\rm Re}(\phi_2^0+\phi^0_1)\right]\, .}\eqn\newtwo$$
Note that this differs from eq.~\new\ due to the new scalar
transformation law [eq.~\transtwo\ instead of eq.~\trans]. From
eq.~\newtwo, we see that the $\ha VV$ vertex
can be generated  even in the  {\it custodial} limit.
Note however that the $\hc W^\mp Z$ vertex is still absent in the same
limit.
When we turn on the U(1)$_Y$ gauge interactions, new trilinear
 $HWB$ and $HBB$ vertices can be generated.\foot{In the case of the
 $\hc W^\mp B$ vertex, the virtual quark-loop contribution does not yield
a term that grows quadratically in the quark mass at
one-loop\refmark\hwz.}
Notice, however, that the
above conclusions are still valid
up to terms of $\calo (m^2_W/E^2)$ where $E$ is the energy
\REF\equite{J.M. Cornwall, D.N. Levin and G. Tiktopoulos, Phys. Rev.
 {\bf D10} (1974) 1145 [E: {\bf D11} (1975) 972];
 B.W. Lee, C. Quigg and H.B. Thacker, Phys. Rev. {\bf D16} (1977) 1519;
M.S. Chanowitz and M.K. Gaillard, Nucl. Phys. {\bf B261} (1985) 379;
Y.P. Yao and C.P. Yuan, Phys. Rev. {\bf D38} (1988) 2237;
J. Bagger and C. Schmidt, Phys. Rev. {\bf D41} (1990) 264.}
of the vector
bosons. This can be seen using the equivalence theorem\refmark\equite\
which states that
the vector bosons can be replaced by their correponding
Goldstone bosons ($G$) in processes with  $E\gg m_W$.
Proceeding as before, it is possible to show that demanding
  {\it custodial} invariance in the quark-Yukawa
interactions, the $HGG$ vertices ($H=\ha ,\hc$ for the case I and
$H=\hc$ for the case II)
are zero.

In order to estimate the contribution of the Higgs self-interactions
and Higgs-gauge interactions to the $HVV$ vertices,\foot{In fact,
gauge and  Higgs loops do not contribute to $\ha VV$ vertices
to all orders in perturbation theory\refmark\kao.}
 we can make use of the
same arguments of the previous section, \ie, these contributions are
expected to be small in the MSSM where the {\it custodial}
SU(2) symmetry is slightly
violated.
Moreover, the contribution of the Higgs-gauge interactions
must vanish in the limit $m_{\hl}=m_{\hh}$.
This can be seen by noting that the kinetic scalar term
 is invariant under the rotation of
the Higgs doublets:
$$\call_{kin}=\sum^2_{i=1}\left(D^\mu\Phi_i\right)^\dagger
\left(D_\mu\Phi_i\right)=
\sum^2_{i=1}\left(D^\mu\Phi^\prime_i\right)^\dagger
\left(D_\mu\Phi^\prime_i\right)\, ,\eqn\alexm$$
where $\Phi^\prime_i$ are defined such as $\VEV{\Phi^\prime_1}=v\equiv
\sqrt{v^2_1+v^2_2}$  and
$\VEV{\Phi^\prime_2}=0$. In the limit $m_{\hl}=m_{\hh}$, we have
$$\Phi^\prime_1=
\left(\matrix{G^+\cr
v+\coeff{1}{\sqrt{2}}\left(\hl +iG^0\right)}\right)\ \ \ \
\Phi^\prime_2=
\left(\matrix{H^+\cr \coeff{1}{\sqrt{2}}\left(
 \hh +i\ha\right) }\right)\, ,\eqn\alexn$$
so that
  $\Phi^\prime_1$ and  $\Phi^\prime_2$ do not mix with each other.
Since  $\VEV{\Phi^\prime_2}=0$, the kinetic term of $\Phi^\prime_2$ is
still SU(2)$_L\times$U(1)$_Y$ invariant after SSB.
Thus, $\Phi^\prime_2VV$
vertices cannot be generated.
\bigskip
\noindent {\bf 4. Conclusions}

We have analyzed the radiative corrections to the charged Higgs mass
in the MSSM and to the $HVV$ vertices by making
use of approximate global symmetries of the theory with two
 Higgs doublets.

In the analysis of the
charged Higgs mass, we have shown that one-loop radiative
corrections from top quarks and top squarks cannot be of
$\calo (g^2m^4_t/m^2_W)$ in the absence of
 $\tilde t_L-\tilde t_R$ mixing.
This has been accomplished by  analyzing the
limit of $g=m_{12}=h_b=0$  where the Higgs potential possesses
 a global  O(4)
$\times$O(4) symmetry. In this limit the charged Higgs boson is a
pseudo-Goldstone  boson associated with the breakdown O(4)$\times$O(4)
$\rightarrow$O(3)$\times$O(3).
By studing the global symmetry properties of the other sectors of the
theory, the dependence of $m^2_{\hc}$ on the model parameters can be
ascertained.

In the analysis of the one-loop effects to the trilinear $HVV$
vertices,
we have shown that a {\it custodial} SU(2) symmetry plays a crucial
role. The appropriate definition
 of the SU(2) symmetry depends
on two possible choices for the pattern of Higgs-fermion couplings.
In the first case, the CP-odd Higgs and charged Higgs
couplings to $VV$ generated at one-loop are zero if the theory is
 {\it custodial} invariant.
In the second case only the charged Higgs couplings to $VV$ are zero in
this limit.
To evaluate the order of magnitude of the radiative corrections to such
$\ha VV$ and  $\hc W^\mp Z$
 vertices, we have
studied the {\it custodial} limit of the different sectors of the theory.
In the MSSM, one learns why in the limit $g^\prime=0$
the $\hc W^\mp Z$ vertex is the only
$HVV$ vertex that does not receive  contributions from
a heavy degenerate fermion doublet. Moreover, due to
the approximate invariance of the Higgs potential and the Higgs-gauge
interactions under the {\it custodial}
SU(2), contributions from the gauge/Higgs
 sectors of the
theory to the $\hc W^\mp Z$ vertex must be very small.
\vskip .7cm
\centerline{\bf Acknowledgements}

We gratefully acknowledge conversations with Marco D\'\i az, Ralf
Hempfling and Graham Ross. This work was supported in part by the U.S.
Department of Energy. The work of A.P. was supported by a fellowship
from the MEC (Spain).
\vfill
\endpage
\refout
\end